\documentclass[aps,prl,twocolumn,nofootinbib]{revtex4}  
\usepackage{graphicx}
\usepackage{amsfonts}
\usepackage{amssymb}
\usepackage{epsfig}
\usepackage{psfrag}
\usepackage{dsfont}
\usepackage{bm} 

\newcommand{\beqn}{\begin{eqnarray}}
\newcommand{\eeqn}{\end{eqnarray}}
\newcommand{\be}{\begin{equation}}
\newcommand{\ee}{\end{equation}}
\newcommand{\GeV}{ {\rm GeV} } 

\usepackage{xcolor}

\begin{document}
\begin{center}
\end{center}

\title{Well-Mixed Dark Matter and the Higgs}

\author{ Daniel Feldman and Pearl Sandick}

\affiliation{Department of Physics and Astronomy, University of Utah, Salt Lake City, UT 84112, USA}

\begin{abstract}
The breaking of electroweak symmetry through renormalization group flow in models 
that have MSSM spectra is found to produce ``well-mixed" neutralino dark matter with a relic density
consistent with the WMAP data and elastic scattering cross section with nuclei consistent with current limits from direct dark matter searches.  These models predict a Higgs boson mass in the range (125-126)~GeV. 
Well-mixed neutralino dark matter is predominantly bino-like, but has significant Higgsino and wino content, each with fractions of comparable size.  With a $\sim1$ TeV gluino mass and sizable neutralino-nucleon scattering cross sections, natural models will be fully tested by both the LHC and future dark matter direct detection experiments.
\end{abstract}

\maketitle 

\noindent{\it  \bf Introduction. $-$} 
The discovery of  a  boson with mass of $\sim125$ GeV \cite{1,2} lends  support for the
the existence of softly broken local supersymmetry. The reason is clear: softly broken local supersymmetry
generally gives rise to a non-vanishing ratio of soft  scalar trilinear to bilinear    
couplings each with mass scaled by the gravitino mass~\cite{Chamseddine:1982jx,Hall,KL,Brig}.  The soft breaking masses and renormalization group (RG) running
then  generate the necessary quantum corrections
to the Higgs mass~\cite{loop}.
Many well motivated models of soft breaking include a Higgs mass that is consistent with the LHC data, e.g.~\cite{Feldman,Akula:2011jx,Djouadi,Akula,Feng,Baer1,Baer2,Shafi,Ellis:2012nv,Yanagida}. The results generally require multi-TeV scalar  {superpartner} masses.
At the mass scale at which electroweak symmetry is broken, naturalness requires that the gaugino-Higgsino sector have suppressed
masses relative to a multi-TeV scale gravitino mass. By ``naturalness'' we simply mean that the Higgsino mass parameter, $\mu$, is not excessively large relative to the mass of the $Z$ boson.  For a recent discussion  on the variable definitions of naturalness for a broad class of models see \cite{Feng:2013pwa}.

The suppression of gaugino masses relative to the scalar superpartner masses means that the lightest supersymmetric particle (LSP) mass is in the range that is being probed by dark matter  experiments.
In particular, neutralino dark matter~\cite{Goldberg,Olive,Drees} remains a leading and viable candidate for particle dark matter.  We will show in this work that neutralino dark matter with natural values of  $\mu$ can lead to a signal  of dark matter in direct detection experiments while yielding the correct relic abundance of cold dark matter in the universe as observed by the WMAP satellite and others~\cite{wmap}, realized within a model with softly broken supersymmetry and REWSB that predicts a mass for the lighter CP-even Higgs boson consistent with that measured at the LHC~\cite{1,2}.

There are several key features of the models discussed here that evade current constraints.  First, at the scale at which the gauge couplings unify (hereafter the unification scale), the  gaugino soft masses are split, i.e.~they are non-universal.  Through renormalization group flow, the  lightest neutralino ($\tilde{N}_1$) mass and the lighter chargino ($\tilde{C}_1$) mass can become nearly degenerate at the electroweak scale,
thus allowing for $\tilde{N}_1 - \tilde{C}_1$ coannihilations \cite{GS,Yamaguchi,Edsjo} in the early universe (coannihilations have recently been revisited in several models~\cite{CoFermi,CoFermi2,HY}) that result in thermal relic neutralino dark matter with the correct abundance. 
Another important feature of the models we discuss is that 
viable neutralino dark matter candidates are a mixture of bino, wino, and Higgsino eigenstates.
The expected neutralino-nucleon elastic scattering cross sections for these models are within reach of current and next generation direct detection experiments, while 
 the continuum gamma-ray flux remains below the current Fermi-LAT sensitivity (for the non-thermal case see \cite{Feldman2,Acharya}). 

We add here that Ref.~\cite {AH} has coined the term ``well-tempered'' neutralino, defined by  $|M_1| \simeq |M_2|$  {\it or}  $|\mu|  \simeq |M_1|$, where $M_1$ and $M_2$ are the electroweak gaugino soft masses.  In the models we discuss here,   $|M_1| \simeq |M_2|$, while {\it also} having $|\mu|  \approx {\rm few} \times |M_1|$ over a significant region of the parameter space.  This results in an LSP that is predominantly bino-like, with a few percent admixture of {\it both} wino and Higgsino components.
We will refer to this arrangement, when the wino the Higgsino fractions are close in value,  as a {\it ``well-mixed''} neutralino.
This model is theoretically well-motivated and gives rise to dark matter and collider signatures within observational reach.
\\

\noindent{\it  \bf Breaking Electroweak Symmetry \& the Higgs. $-$} 
Recently an interesting part of the supergravity parameter space has been uncovered \cite{Feldman} where
the square of the soft mass for the up-type Higgs runs small and positive under RG flow leading
to the breaking of electroweak symmetry with a rather low value of the $\mu$ term for heavy soft breaking scalars at a mass scale of $\sim10$'s of TeV~\cite{Feldman,Akula:2011jx,Baer2}.  As noted in Ref.~\cite{Feldman}, the result is not a focus point solution, but instead a new solution to electroweak symmetry breaking owing to the  cancellation of RG parameters defined at the unification scale.  

To see how the cancellation works, one need only examine the running square of the soft mass for the up-type Higgs, $M_{H_u}^2(t)$, where $t = \ln(Q/Q_0)$ with $Q$ and $Q_0$ denoting the energy scale and the unification scale, respectively. The soft breaking mass for the up-type Higgs can be written in terms of RG-dependent functions $r_i(t)$ and the soft breaking masses and couplings for the scalars and gauginos. In the one loop approximation the RG equations (RGEs) can be solved analytically giving rise to 
\beqn 
M^2_{H_u}(t) &=&  r_1(t) M^2_0 - r_2(t)  A^2_0 + \epsilon(t) \label{eq:mhusq} \\  \epsilon(t) &=& r_3(t) A_0 M_{a} +  r_4(t) M^2_{a}+\ldots
 \label{XX}
\eeqn
where $M_0$ and $A_0$ are the universal scalar soft masses and scalar trilinear couplings, and $M_{a}$ are the gaugino masses, with $a ={1,2,3}$ for $SU(3)$, $SU(2)$, and $U(1)$ respectively, all defined at the unification scale $Q_0 \approx 2\times10^{16}~\GeV$.
For the case of heavy scalars with suppressed gaugino masses, the
 term $\epsilon(t)$ is a residual correction and is small. The  coefficients of $M_0$ and $A_0$ at one loop are
$r_1(t) = \frac{1}{2}(3 \delta(t) -1)$ and $r_2(t) =\frac{1}{2}( \delta(t) - \delta^2(t))$,
where $\delta(t)$ depends on the gauge couplings and  on the top Yukawa. 
As found in Ref.~\cite{Feldman}, for electroweak symmetry breaking triggered by a heavy stop, i.e.~$Q_{\rm EWSB} \equiv Q^*$ 
where $Q^*= \sqrt{ m_{\tilde t_1} m_{\tilde t_2} }$, the RG functions $r_1(t)$ and $r_2(t)$ begin to approach a common positive value
\beqn
r_1(Q^*) \simeq r_2(Q^*) \sim {\cal O}(1/10).
\eeqn
This phenomenon has been referred to as an intersection point (IP) \cite{Feldman,Douglas} of the RG flow since the first two terms on the right hand side of Eq.~(\ref{eq:mhusq}) ``intersect'' and can cancel. 

The IP presents the opportunity to drastically reduce $M^2_{H_u}$ relative to $M^2_0$ and $A^2_0$.
In order to achieve the cancellation, it is obvious that $r_1(Q^*)M_0^2$ and $r_2(Q^*)A^2_0$ should be nearly degenerate. Since $r_1(t) \simeq r_2(t)$, the cancellation requirement becomes a statement that the ratio of the soft parameters $|A_0|/M_0$  approaches unity.  We note that a  shift in the top pole mass will shift a particular IP value of $|A_0|/M_0$, however the ratio will still be close to unity. 

The relationship between $M_0$ and $A_0$ can be viewed as a direct consequence of string moduli supersymmetry breaking~\cite{KL,Brig}, in which the scalar masses and trilinear couplings  are related to the gravitino mass, $M_{3/2}$, via
\begin{equation}
M_{\alpha}^2 \simeq M_{3/2}^2 \simeq M^2_0
\end{equation}
and
\begin{equation}
A_{\alpha\beta\gamma} \simeq 
F^M \left({\hat K}_M + \partial_M \log Y_{\alpha\beta\gamma} \right)\ \simeq M_{3/2} \simeq A_0,
\end{equation}
where $F^M$ is the order parameter of supersymmetry breaking for moduli $(M)$,  $\hat{K}_M$ is the derivative of the K\"{a}hler potential, and $Y_{\alpha\beta\gamma}$ are Yukawa couplings.
Since $M_0$ and $|A_0|$ are both equal to $M_{3/2}$, up to small corrections, $|A_0|/M_0 \approx 1$.  
Furthermore, the bilinear coupling $B$ for the Higgs sector is consistent with $B_0  \simeq 2 M_{3/2}$. 
At the EWSB scale $Q^*$, $\mu$ is suppressed simply because the IP results in a small value for $M^2_{H_u}$ (which is further suppressed by tadpole corrections).
The down-type Higgs soft mass squared, $M^2_{H_d}$, runs very little, taking a value $\sim M^2_0 \simeq M^2_{3/2}$.
In the parameter space where $\tan \beta \approx  2/\sin 2 \beta$ and where the minimization of the Higgs potential breaks the electroweak symmetry, the value of $\mu$ can be as small as \cite{Feldman}
\be
\mu \approx \frac{ M_{3/2}}{2\tan \beta},
\ee
again, up to small corrections.  This result of a large gravitino mass and  $\mu/M_{3/2}$ being suppressed by the inverse of  $\tan \beta$ has also recently been discussed in Ref.~\cite{Yanagida}.

The determination of the $\mu$ parameter is intimately tied to the mass of the Higgs boson through electroweak symmetry breaking. 
At an intersection point, the light CP even Higgs has a mass near 
\be 
m_{ {Higgs} } = (125 -126)~\GeV.
\ee
We stress that this is a generic prediction of an IP of RG flow, since the sfermion masses must be large, ${\cal O}(10) \rm TeV$, and thus so is $A_0$.
Indeed, the loop correction for the Higgs mass is naturally of the right size. 
This is a consequence of the top trilinear coupling at the EWSB scale, $A_t$, 
and the geometric mean of the stop masses, \mbox{$M_S = \sqrt{m_{\tilde{t}_1} m_{\tilde{t}_2}}=Q^*$}, entering the leading loop correction as
 \beqn  X_t /M_S &= &(A_t - \mu/\tan \beta)/M_S   \sim A_t/M_S.  \eeqn
As $A_0/M_0$ runs to $A_t/M_S$ at the electroweak scale, the ratio remains of order unity.  This, along with the relatively large value of $M_S$ as controlled by the RG running, gives  the necessary correction to the light CP even Higgs mass~\cite{Feldman}.

Having addressed the scalar sector masses and dynamics, we turn, finally, to the gauginos.  Suppression of gaugino masses can arise from moduli dominated supersymmetry breaking.
This feature was realized early on in the context of string model building~\cite{Ibanez} where the moduli contribution to supersymmetry breaking can dominate over the dilaton contribution.

More generally, in Planck units the gravitino mass is
$M^2_{3/2} =  \frac{1}{3} \langle {\bar F}^{\bar I}{\hat K}_{{\bar I}J} F^J \rangle $ so the gravitino can become massive
via the Super-Higgs mechanism with a single dominant $F$-term and other $F$-terms suppressed.  
At the unification scale,  tree and loop contributions to the gaugino masses
 can have comparable sizes since the modulus that supplies the dominant $F$-term will lead to a loop-suppressed contribution to the gaugino masses (see e.g. \cite{NIR} for a pedagogical analysis). 
 Thus, the  gaugino masses will be suppressed relative to the gravitino mass and hence relative to the scalar superpartner masses,
\be
M_a(Q_0)  =  {\cal{O}}_a(10^{-2}) \cdot M_{3/2}.
\ee
Note that the soft masses and couplings for the scalar sector of the theory, $M_0$, $A_0$, and $B_0$, are still dominated by an unsuppressed $|F^M| \sim M_{3/2}$.
The precise ratios of the gaugino masses at the unification scale are, of course, model dependent. 
\\

\noindent{\it  \bf Well-Mixed Neutralino. $-$} 
For simplicity, we consider $M_0 = 10 ~\rm TeV$, $A_0/M_0 = -0.75$,  $\tan \beta = 10$, and $m_t(\text{pole})$ = 173~GeV, and use SOFTSUSY~\cite{SS} for the renormalization group flow and micrOMEGAs~\cite{micrOMEGAs} to calculate the annihilation cross sections and spin independent scattering cross sections, varying the gaugino masses at the unification scale, $M_{1,2}(Q_0)$.  We allow $150 ~{\rm GeV} \leq M_{1,2}(Q_0) \leq 2 ~{\rm TeV}$, and $M_3(Q_0) \geq 300 ~\rm GeV$, as below this lower limit the gluino mass is $M_{\tilde{g}} \lesssim 850$ GeV and is constrained by the LHC. Naturalness generally requires  $M_3(Q_0) \lesssim 500 ~\rm GeV$, however we consider $M_{\tilde g}$ up to 1.8 TeV.  Given these parameter ranges, the neutralino LSP mass is always $\gtrsim 100$ GeV, thereby avoiding the $Z$ boson and light CP even Higgs poles (for a recent dedicated analysis of the pole regions, see \cite{Freese}).

\begin{figure}[h]
\vspace*{.2in}
\hspace*{-.2in}
\centering
\includegraphics[width=9cm,height=7cm]{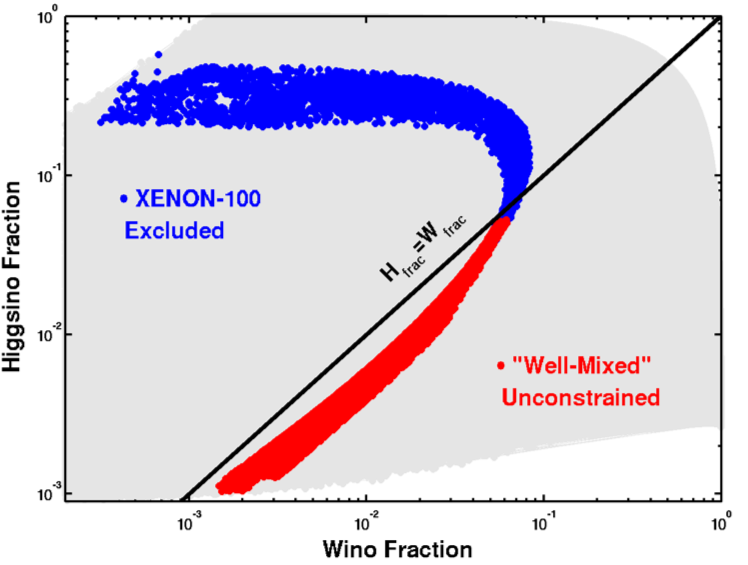}
\caption{ Wino fraction vs. Higgsino fraction:  Grey points have successful REWSB, but no dark matter constraints are implemented.  Blue and red points have thermal relic neutralino dark matter with the correct abundance.  Red points are consistent with the current limits from direct dark matter searches (i.e. the current XENON-100 limit).
}
\label{fig:money}
\end{figure}

The eigencontent of the neutralino LSP is represented as $Z_{1i}$, where $i=1,2,(3,4)$ for the bino, wino, and (two Higgsino) component(s), respectively, i.e.~$\tilde N_1  = Z_{11} \tilde B + Z_{12} \tilde W + Z_{13} \tilde H_1 + Z_{14} \tilde H_2$. 
Defining the Higgsino fraction as   $H_F =\left| Z_{13}\right|^2 + \left|Z_{14}\right|^2,$ and wino fraction as  $W_F =|Z_{12}|^2 $, we plot in   Fig.~\ref{fig:money} ${H_F}$ vs. ${W_F}$ for each model point.  We note that this figure includes only neutralino masses up to 500 GeV, as those will be the subject of the analysis in the following sections, though extending to larger neutralino masses leaves the picture essentially unchanged.

A quick inspection of Fig.~\ref{fig:money} reveals that essentially all unconstrained models have ${H_F} \simeq {W_F}$; these are the ``well-mixed" models.  In the absence of any dark matter constraints, all neutralino compositions are allowed (light grey points).  Once the constraint on the relic abundance of neutralinos is imposed, only the red and blue points survive.  A large wino fraction leads to a dearth of neutralino dark matter relative to the measured value, thus we see that all red and blue points have small wino fractions, i.e ${W_F} \lesssim 0.06$.  Points with large Higgsino fractions, however, may be compatible with the dark matter abundance.  In addition to the relic abundance of neutralino dark matter, we apply the constraint on the spin independent neutralino-nucleon elastic scattering cross section from XENON-100~\cite{xenon}, evaded only by red points.  Model points with large Higgsino fractions have large scattering cross sections with nuclei, thus all points consistent with the XENON-100 bound have Higgsino fractions $H_F \simeq W_F$, i.e.~${H_F} \lesssim 0.06$.
We note that the spin independent scattering cross section does not, by itself, constrain the wino fraction.  When combined with the relic abundance constraint, however, dark matter direct detection experiments will probe the largest wino/Higgsino fractions down to the smallest in these models.  We see that all points that survive both the constraint on the relic abundance of neutralino dark matter and that on the spin independent neutralino-nucleon elastic scattering cross section have approximately comparable wino and Higgsino fractions, and are therefore ``well-mixed''.

\begin{table}[t]
\begin{center}
\begin{tabular}{|c|c|c|c|c|c|c|c|} 
\hline
Model & A & B & C \\
\hline\hline
$M_1$   {\rm [GeV]} & 300   &290 & 370 \\
$M_2$& 169  &  167 & 210 \\
$M_3$& 300& 400 & 440 \\
\hline
$M_0$  {\rm [TeV]} & 10   & 10      & 10  \\
$\tan \beta$&  10&  13 &  12 \\
$A_0/M_0$ &  -0.75& -0.74 & -0.72 \\
$m_t(\text{pole})$ {\rm  [GeV]} &  173  & 173 & 173 \\
  \hline
 $Z_{11}$         &                0.974  &   0.962 & 0.965 \\
 $Z_{12}$         &             -0.182  & -0.211 & -0.206\\
$Z_{13}$   &                0.126 & 0.162 & 0.150\\
$Z_{14}$     &             -0.046 & -0.062 & -0.065 \\
\hline
 $\sigma_{SI}$  [pb] &  1.2 $\times 10^{-9}$  & 2.0 $\times 10^{-9}$  & 2.2 $\times 10^{-9}$ \\
 \hline
$m_{Higgs}$         [GeV]   & 126  & 126   & 126 \\
$M_{\tilde N_1} $    &  134  &  128   & 164\\
$M_{\tilde C_1} $     & 153      & 148    & 185\\
$M_{\tilde N_2} $    & 154 &  149   &  186\\ 
$M_{\tilde N_3} $   &  506  &  419   & 465 \\
$M_{\tilde N_4} $   &  515  & 430   & 477 \\
$M_{\tilde C_2} $    &  516 & 431  & 477 \\
$M_{\tilde  g} $     &   860 & 1107   & 1201\\
$\mu$   &486 &  400   & 446 \\ 
\hline
  $\Omega h^2$ &    0.12 &    0.12 &    0.11 \\
 \hline
\end{tabular}
\caption{Sample models with well-mixed neutralino dark matter.
}    
\label{tabbench}
\end{center}
\end{table}

The cosmologically-preferred relic abundance of well-mixed neutralinos is achieved through coannihilations of the neutralino LSP with the slightly heavier neutralinos and charginos in conjunction with an advantageous bino-wino-Higgsino LSP eigencontent.    Recall that  the relic density of neutralinos is inversely proportional to the effective annihilation cross section in the early universe, the latter  depending on Boltzmann factors and spin degrees of freedom of the states that coannihilate. The mass dependencies appear in the Boltzmann factors, and so are exponentially suppressed for coannihilating states; 
${\rm exp}(- \Delta_p M_{N_1}/T)$, where $T$ is the temperature and 
\be 
\Delta_p = M^{-1}_{\tilde N_1} ( M_{\tilde p} -M_{\tilde N_1}), ~{\rm with} ~ p= \{\tilde C_1,\tilde N_2\}. \label{eq:delta}
\ee
Over the well-mixed neutralino parameter space (red points in Fig.~\ref{fig:money}), the contribution of $\tilde N_1 \tilde N_1 \to WW$
to the relic density is typically no more than $\sim20 \%$ with coannihilations therefore contributing dominantly for $\Delta_p \sim 0.05-0.15$ with smaller neutralino mass corresponding to a larger $\Delta_p$.  
The analysis allows for arbitrary ratios of $M_1/M_2$ at the unification scale,  however a majority of the well-mixed models congregate in the region $M_1/M_2 \in (1.6 - 1.9)$.

In Table~\ref{tabbench} we present three sample model points that demonstrate the particle spectra and dark matter observables explicitly.
Sample models B and C arise from small perturbations about the parameter space sweep, resulting in a similar suite of dark matter observables.
It is evident that movement away from our fixed parameter choices, i.e.~$A_0/m_0$, $m_t(\text{pole})$, and $\tan\beta$, can result in small shifts in the value of $\mu$, which can change, somewhat, the spin independent scattering cross section, as evident in Table \ref{tabbench}.  Nonetheless, the neutralino LSP remains well-mixed, and our conclusions are robust.
The light CP even Higgs mass is about 126 GeV in each model.  \\

\noindent{\it  \bf Implications for Direct \& Indirect Detection. $-$}
At an intersection point \cite{Feldman}, the  gaugino masses are suppressed relative to the scalar masses, as noted below~Eq.~(\ref{XX}).  The fact  that $\mu$ takes on a value as low as $\sim$ few$\times M_Z$, comparable in size to $M_1$ and $M_2$, leads to a well-mixed neutralino LSP. The resulting bino-wino-Higgsino mixings have an important impact on dark matter observables, as has been studied for different models of soft breaking \cite{PS}, as well as on LHC signals of the resultant spectrum~\cite{CD}.

\begin{figure*}[t]
\vspace*{.2in}
\hspace*{-.2in}
\centering
\includegraphics[width=9cm,height=7cm]{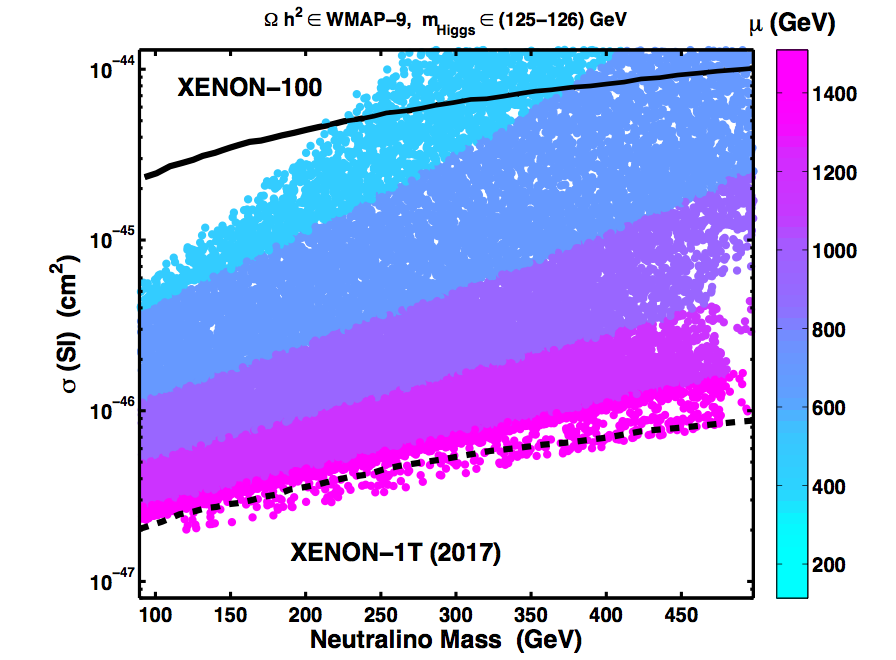} \hspace{5mm}
\includegraphics[width=8.3cm,height=7cm]{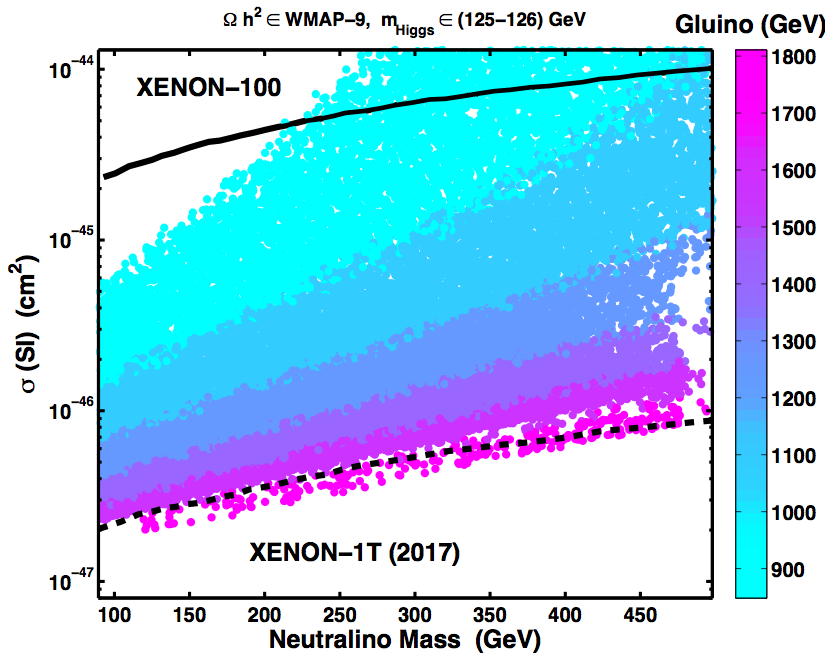}
\caption{The spin independent neutralino-nucleon elastic scattering cross section is plotted as a function of neutralino mass, with points color-coded by the value of $\mu$ (left panel) and the gluino mass $M_{\tilde g}$ (right panel). The relic density of cold dark matter lies within the WMAP band for all model points, and the Higgs mass is in the range (125-126) GeV. Also shown is the present XENON-100 limit~\cite{xenon} and the XENON-1T projected sensitivity~\cite{xenon1T}.  }
\label{data3} 
\end{figure*}

In Fig.~\ref{data3} we present the neutralino-nucleon spin independent elastic scattering cross section as a function of neutralino LSP mass for points that yield a relic abundance of neutralino dark matter in the cosmologically-preferred range.  Points are color-coded by the value of $\mu$ in the left panel and the gluino mass, $M_{\tilde{g}}$, in the right panel.  The correlation between $\mu$ and $M_{\tilde{g}}$ is clear.  
Had we considered $M_{\tilde{g}} \gtrsim 1.8$ TeV, there would be viable models with cross sections below the XENON-1T projection, however naturalness arguments point to small $\mu$ and therefore small $M_{\tilde{g}}$.  

The models presented here have MSSM scalars so heavy that they are effectively decoupled and therefore contribute minimally to the spin independent neutralino-nucleon elastic scattering cross section, $\sigma_{SI}$.  In the limit of very heavy scalars, $\sigma_{SI}$ takes on a simple analytical form (see e.g. \cite{hoop} for an overview) where the dominant contribution arises from $t$-channel exchange of the light CP-even Higgs. Though we acknowledge that hadronic uncertainties are important for a precise determination of $\sigma_{SI}$~\cite{Savage}, here we take the default  values for the hadronic matrix elements and the pion-nucleon sigma term, $\Sigma_{\pi N}$, as in \cite{micrOMEGAs}. 
In Fig.~\ref{data3}, we see that these models can yield $\sigma_{SI}\approx 10^{-44} ~\rm cm^2$ for neutralino masses  above  $ \sim 250 ~\rm GeV$, already constrained by current experiments. Most models presented here have cross sections that will be probed by next generation direct dark matter searches such as XENON-1T~\cite{xenon1T}, LUX~\cite{LUX}, and SuperCDMS~\cite{supercdms}, and models with increasing $\mu$ (increasing $M_{\tilde{g}}$) will be subsequently tested as experimental sensitivities improve.   

\begin{figure}[h,t]
\vspace*{.2in}
\hspace*{-.2in}
\centering
\includegraphics[width=9cm,height=7cm]{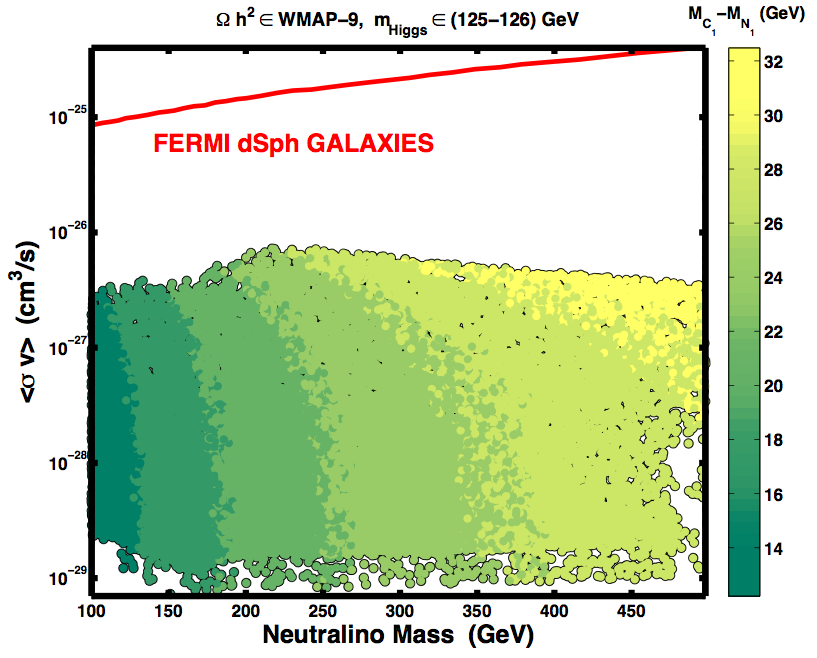}
\caption{ Cross sections to continuum photons $\langle \sigma v \rangle$ for annihilations at the current temperature. The models shown satisfy the XENON-100 constraint. The cross sections are dominated by $WW$ final states for well-mixed dark matter. Also shown are the limits from the combined analysis of dSph Milky Way satellites~\cite{Ackermann:2011wa}.  Points are shaded by the mass splitting between the neutralino LSP and lighter chargino.
The relic density of cold dark matter lies within the WMAP band for all model points, and the Higgs mass is in the range (125-126) GeV. 
}
\label{data2}
\end{figure}

In  Fig.~\ref{data2} we show the annihilation cross section to continuum photons for model points that pass the dark matter abundance and direct detection constraints (the red points in Fig.~\ref{fig:money}).  For comparison, we also show the constraints derived from a combined analysis of dwarf spheroidal (dSph) Milky Way satellite galaxies~\cite{Ackermann:2011wa,GeringerSameth:2011iw}. 
Milky Way dSphs are excellent targets for dark matter searches; they are dark matter dominated objects that contain few, if any, sources of gamma-ray photons that would constitute a background to a dark matter annihilation signal (for extended discussions see \cite{GeringerSameth:2011iw,R}).  Ref. ~\cite{Hooper:2012sr} has also explored bounds on generic models of dark matter annihilation.  For the models we discuss here, the bounds from \cite{Hooper:2012sr} are comparable to those from the gamma-ray flux from Milky Way dSphs~\cite{Ackermann:2011wa,GeringerSameth:2011iw}.

The continuum photon spectrum from dark matter annihilations today comes almost exclusively from $\tilde N_1 \tilde N_1 \to WW$ and is well known to be enhanced
over different regions of the parameter space (for example, for a pure wino). This is not the case for well-mixed dark matter.
  The  $ \tilde N_1 \tilde N_1 \to WW$ amplitude  (see e.g.~\cite{Drees,report}) depends on $O^{L}_{1j} =  -\frac{1}{\sqrt 2} Z_{14} V_{j2} +Z_{12} V_{j1}$ and 
$O^{R}_{1j} =  +\frac{1}{\sqrt 2} Z_{13} U_{j2} +Z_{12} U_{j1}$,
where $U$ and $V$ are the mass matrices that diagonalize the chargino sector.
The amplitude for annihilation to $WW$, ${\cal A}_{WW}$, that arises in the galactic halo is proportional to the  products $O^{L}_{1j} O^{L}_{1j}$ and $O^{R}_{1j} O^{R}_{1j}$. Thus the cross section for  $WW$ is proportional to the neutralino and chargino eigencomponents  to the fourth power.  We also add that the effects of bremsstrahlung are not too large \cite{brem,brem2} as the mass of the neutralino is constrained by naturalness and the kinematic endpoint of the $W$ fragmentation distribution is limited by phase space. In terms of the line cross sections for the production of photons, we have verified with DarkSusy~\cite{DS} that the line cross sections are suppressed at the level of $ \sim 10^{-29}~\rm cm^3/s$ or smaller over all models. This is several orders of magnitude smaller than the constraints from the Fermi line searches~\cite{B,A}.

In Fig.~\ref{data2}, the color of each model point indicates mass splitting between the well-mixed neutralino dark matter and the lighter chargino.  
The mass splittings shown  have important consequences for LHC searches.\\

\noindent{\it  \bf  Gauginos at the LHC $-$} 
Summarizing the relative mass scales in the model, we have:
\begin{eqnarray} 
\{M_1,M_2,M_3, \mu \} \ll M_{3/2} \nonumber \\
 M_0 \sim |A_0| \sim B_0 \sim M_{3/2}.
 \label{softy}
 \end{eqnarray}
The {\it observable LHC spectra}, i.e.~particles conceivably light enough to produce 
a significant number of events above the Standard Model background, are the light CP-even Higgs, four neutralinos, two charginos, and the gluino. Of immediate relevance to the LHC,
\begin{eqnarray}
m_{Higgs} & \sim& 126 ~\GeV, \nonumber \\
M_{\tilde g} & \sim&  ~(1-1.5)~{\rm TeV}, \nonumber \\
M_{\tilde N_1} & \sim&  {\cal O}(10^2~{\rm GeV}), \\
M_{ \tilde C_1}-M_{\tilde N_1} & \sim&  (10-30)~ {\rm GeV}, \nonumber 
\end{eqnarray}
with $M_{ \tilde C_1} \simeq M_{\tilde N_2}$.
The fact that the electroweak sector gaugino masses are $\sim \mathcal{O} (100$ GeV), and that the gluino is in the TeV range for $\mu$ of natural size implies that the discovery of new Majorana fermion states should be possible at the LHC. 

The CMS and ATLAS searches \cite{gauginosLHC} are not yet sensitive to neutralino masses of $100-500$ GeV for 
mass splittings as small as those indicated in Fig.~\ref{data2}.
CERN's proposed compact linear collider (CLIC)~\cite {CLIC} or the proposed International Linear Collider (ILC)~\cite{ilc} would, however, be able to
resolve the mass splittings in the gaugino sector.
The upgraded LHC will certainly test whether the scenario described here provides an adequate description of nature:
The first aspect of this scenario to be tested by the LHC will be the prediction of a relatively light gluino.  Currently, a \mbox{$\sim \mathcal{O} (1$ TeV)} gluino with decoupled squarks is  consistent with the LHC data (see e.g.~\cite{deJong}), with the precise gluino mass constraint being model dependent and strongly correlated with the branching ratios of the gluino. The maximal gluino mass 
considered here is $\sim 1.8~ \rm TeV $, as controlled by the prejudice for naturalness and REWSB. 
The present (2013-2014) upgrade to the LHC will give it the capability to probe this entire model class (for a recent re-analysis of LHC reach in gluino mass at larger center of mass energy, see \cite{review,BBLT}), with LHC gluino mass reach being complementary to the sensitivity of direct dark matter searches to neutralino-nucleon scattering, as demonstrated in the right panel of Fig.~\ref{data3}.

The  potential for discovery of these new states by the LHC is limited by three factors: (1) the previous center of mass operating energy and luminosity to date, (2) the small mass splittings of the LSP and the lighter chargino and second lightest neutralino restrict the phase space of the chargino decay, and (3) the amount of missing energy  produced in the three-body decays of the gluino lead to an effective mass  {(sum of the jet transverse momenta and missing energy)}  distribution, including the  peak,  that is similar to that expected from the Standard Model background.  Observation of events from gluino and electroweak gaugino cascades in these models would require more LHC data. Currently, the very lowest gluino masses predicted in this model class are likely constrained by the LHC (though, strictly speaking, a dedicated analysis is required), and in some cases also by direct dark matter searches.  
The upgraded LHC will likely have the capability to test this entire model class for natural values of $\mu$ with just a few years of data.
\\

\noindent{\it  \bf  Conclusions. $-$} 
This work demonstrates that well-mixed neutralinos are viable dark matter candidates arising in well-motivated models that predict natural values of the $\mu$ parameter
and a Higgs mass that takes the observed value of the new boson near 126 GeV. The annihilation cross section to continuum photons and resultant gamma-ray flux from annihilation in the galaxy are unconstrained by indirect detection experiments. Direct detection experiments are beginning to constrain well-mixed dark matter, and next generation detectors will be sensitive to nearly all of the models explored in this study.

The parameter space of well-mixed dark matter derives from the radiative breaking of electroweak symmetry, where the $\mu$ parameter and physical gaugino masses are suppressed relative to the heavy scalar superpartners.  The relic density is satisfied largely by coannihilations in the early universe.  The gluino is in the TeV domain and will produce multi-jets accessible to future runs at the LHC. Central to this conclusion is the fact that the unification-scale soft breaking scalar masses and couplings are all of order the gravitino mass (see Eq. \ref{softy}), which is a  generic prediction of supergravity and string-motivated models of soft breaking.

Within this framework, desirable features of supersymmetry remain intact.
The soft breaking of supersymmetry incorporates gravity via the gauging of global supersymmetry,
and the mass generation for superpartners occurs via the Super-Higgs effect, breaking supersymmetry and thus  generating
the soft masses. The models exhibit gauge coupling unification  and dynamically trigger spontaneous
electroweak symmetry  breaking through renormalization group flow.
The dark matter candidate is the long-coveted neutralino, 
which will have a mass near the electroweak scale.
The model is  predictive - unification scale boundary conditions determine TeV scale phenomena - and the most natural regions of parameter space are fully testable with the LHC and with dark matter direct detection experiments.  Well-mixed neutralino dark matter may be just around the corner.  
\\
\\\noindent{\it  \bf  Note Added and Acknowledgments $-$} 
 We would like to thank Matt Reece, Neal Weiner,
Sasha Pukhov and Genevi\`{e}ve  B\'{e}langer for helping to uncover and interpret 
an error in the photon line cross section calculation in micrOMEGAs V2.4.5 (wherein the tree level neutralino mass matrix is used to produce the photon line cross section) used in an earlier version of this paper. 
D.F.~would like to thank the Department of Physics and Astronomy at the University of Utah for support and hospitality throughout the completion of this work.  Support and resources from the Center for High Performance Computing at the University of Utah are also gratefully acknowledged.


\vspace{-.2cm}
\begin{thebibliography}{999}

\vspace{-.15cm}
\bibitem{1} 
  G.~Aad {\it et al.}  [ATLAS],
  Phys.\ Lett.\ B {\bf 716}, 1 (2012).


\bibitem{2} 
  S.~Chatrchyan {\it et al.}  [CMS],
  Phys.\ Lett.\ B {\bf 716}, 30 (2012).




  
\bibitem{Chamseddine:1982jx} 
  A.~H.~Chamseddine, R.~L.~Arnowitt and P.~Nath,
  Phys.\ Rev.\ Lett.\  {\bf 49}, 970 (1982).
  
\bibitem{Hall} 
  L.~J.~Hall, J.~D.~Lykken and S.~Weinberg,
  Phys.\ Rev.\ D {\bf 27}, 2359 (1983).
  
  
\bibitem{KL} 
  V.~S.~Kaplunovsky and J.~Louis,
  Phys.\ Lett.\ B {\bf 306}, 269 (1993);
  Nucl.\ Phys.\ B {\bf 422}, 57 (1994).

  
  \bibitem{Brig} 
  A.~Brignole, L.~E.~Ibanez and C.~Munoz,
  Nucl.\ Phys.\ B {\bf 422}, 125 (1994).


\bibitem{loop} 
 J.~R.~Ellis, G.~Ridolfi and F.~Zwirner,
  Phys.\ Lett.\ B {\bf 257}, 83 (1991);
  H.~E.~Haber and R.~Hempfling,
  Phys.\ Rev.\ Lett.\  {\bf 66}, 1815 (1991);
  Y.~Okada, M.~Yamaguchi and T.~Yanagida,
  Prog.\ Theor.\ Phys.\  {\bf 85}, 1 (1991).

\bibitem{Feldman}
 D.~Feldman, G.~Kane, E.~Kuflik and R.~Lu,
 Phys.\ Lett.\  B {\bf 704}, 56 (2011);
 [arXiv:1105.3765 [hep-ph]].
  B.~S.~Acharya, G.~Kane and P.~Kumar,
  Int.\ J.\ Mod.\ Phys.\ A {\bf 27}, 1230012 (2012);
  G.~Kane, R.~Lu and B.~Zheng,
  arXiv:1211.2231 [hep-ph].
 
 
 
\bibitem{Akula:2011jx} 
  S.~Akula, M.~Liu, P.~Nath and G.~Peim,
  Phys.\ Lett.\ B {\bf 709}, 192 (2012).
 
 
\bibitem{Djouadi} 
  A.~Arbey, M.~Battaglia, A.~Djouadi, F.~Mahmoudi and J.~Quevillon,
  Phys.\ Lett.\ B {\bf 708}, 162 (2012).

\bibitem{Akula} 
  S.~Akula,  B.~Altunkaynak, D.~Feldman, P.~Nath and G.~Peim,
  Phys.\ Rev.\ D {\bf 85}, 075001 (2012).
  
\bibitem{Baer1} 
  H.~Baer, V.~Barger and A.~Mustafayev,
  Phys.\ Rev.\ D {\bf 85}, 075010 (2012).


\bibitem{Feng} 
  J.~L.~Feng and D.~Sanford,
  Phys.\ Rev.\ D {\bf 86}, 055015 (2012).
  
  \bibitem{Baer2} 
  H.~Baer,  V.~Barger, P.~Huang, A.~Mustafayev and X.~Tata,
  Phys.\ Rev.\ Lett.\  {\bf 109}, 161802 (2012).

\bibitem{Shafi} 
  I.~Gogoladze, F.~Nasir and Q.~Shafi,
  arXiv:1212.2593 [hep-ph].

\bibitem{Ellis:2012nv} 
  J.~Ellis, F.~Luo, K.~A.~Olive and P.~Sandick,
  arXiv:1212.4476 [hep-ph].
  
\bibitem{Yanagida} 
  J.~L.~Evans, M.~Ibe, K.~A.~Olive and T.~T.~Yanagida,
  arXiv:1302.5346 [hep-ph].
  
  \bibitem{Feng:2013pwa} 
  J.~L.~Feng,
  arXiv:1302.6587 [hep-ph].
  
\bibitem{Goldberg} 
  H.~Goldberg,
  Phys.\ Rev.\ Lett.\  {\bf 50}, 1419 (1983).
  

\bibitem{Olive} 
  J.~R.~Ellis, J.~S.~Hagelin, D.~V.~Nanopoulos, K.~A.~Olive and M.~Srednicki,
  Nucl.\ Phys.\ B {\bf 238}, 453 (1984).
  
  \bibitem{Drees} 
  M.~Drees and M.~M.~Nojiri,
  Phys.\ Rev.\ D {\bf 47}, 376 (1993).
  
\bibitem{report} 
For a review see:   G.~Jungman, M.~Kamionkowski and K.~Griest,
  Phys.\ Rept.\  {\bf 267}, 195 (1996).
  
  \bibitem{B} 
  M.~Ackermann {\it et al.}  [FERMI-LAT],
  Phys.\ Rev.\ D {\bf 86}, 022002 (2012)
  A.~A.~Abdo, i {\it et al.},
  Phys.\ Rev.\ Lett.\  {\bf 104}, 091302 (2010).
  
  
  
    \bibitem{A}
\url{http://fermi.gsfc.nasa.gov/science/mtgs/symposia/2012}
     
 
  
  \bibitem{wmap}   G.~Hinshaw, {\it et al.}  [WMAP],
  arXiv:1212.5226 [astro-ph.CO].
  
\bibitem{xenon} 
  E.~Aprile {\it et al.}  [XENON100],
  Phys.\ Rev.\ Lett.\  {\bf 109}, 181301 (2012).
  



\bibitem{GS} 
  K.~Griest and D.~Seckel,
  Phys.\ Rev.\ D {\bf 43}, 3191 (1991).
  \bibitem{Yamaguchi} 
    S.~Mizuta and M.~Yamaguchi,
  Phys.\ Lett.\ B {\bf 298}, 120 (1993).
  
  \bibitem{Edsjo} 
  J.~Edsjo and P.~Gondolo,
  Phys.\ Rev.\ D {\bf 56}, 1879 (1997).



\bibitem{CoFermi} 
  D.~Feldman, Z.~Liu, P.~Nath and B.~D.~Nelson,
  Phys.\ Rev.\ D {\bf 80}, 075001 (2009);
    D.~Feldman,
  Nucl.\ Phys.\ Proc.\ Suppl.\  {\bf 200-202}, 82 (2010);
  N.~Chen, {\it et. al},
  Phys.\ Rev.\ D {\bf 83}, 035005 (2011);
  Phys.\ Rev.\ D {\bf 83}, 023506 (2011);
   D.~Feldman and G.~Kane,
  In *Kane, G.L. (ed.): Perspectives on supersymmetry II* 288-304.
  
    \bibitem{CoFermi2} 
  G.~Belanger, C.~Boehm, M.~Cirelli, J.~Da Silva and A.~Pukhov,
  JCAP {\bf 1211}, 028 (2012).
  
  \bibitem{HY}
  S.~Tulin, H.~-B.~Yu and K.~M.~Zurek,
  arXiv:1208.0009 [hep-ph].




    \bibitem{Feldman2} 
  D.~Feldman, G.~Kane, R.~Lu and B.~D.~Nelson,
  Phys.\ Lett.\ B\ {\bf 687}, 363  (2010)
  [arXiv:1002.2430 [hep-ph]].
      \bibitem{Acharya} 
    B.~S.~Acharya, G.~Kane, P.~Kumar, R.~Lu and B.~Zheng,
  arXiv:1205.5789 [hep-ph].

\bibitem{AH} 
  N.~Arkani-Hamed, A.~Delgado and G.~F.~Giudice,
  Nucl.\ Phys.\ B {\bf 741}, 108 (2006).
  

\bibitem{Douglas}
 M.~R.~Douglas,
  arXiv:1204.6626 [hep-th].
  

\bibitem{Ibanez} 
  L.~E.~Ibanez and H.~P.~Nilles,
  Phys.\ Lett.\ B {\bf 169}, 354 (1986);
  A.~Font, L.~E.~Ibanez, D.~Lust and F.~Quevedo,
  Phys.\ Lett.\ B {\bf 245}, 401 (1990);
    L.~E.~Ibanez and D.~Lust,
  Nucl.\ Phys.\ B {\bf 382}, 305 (1992);
  B.~de Carlos, J.~A.~Casas and C.~Munoz,
  Phys.\ Lett.\ B {\bf 299}, 234 (1993).\\
For a compendium of results see:  K.~Choi and H.~P.~Nilles,
  JHEP {\bf 0704}, 006 (2007).

\bibitem{NIR} 
  J.~Louis and Y.~Nir,
  Nucl.\ Phys.\ B {\bf 447}, 18 (1995).

\bibitem{Freese} 
  D.~Feldman, K.~Freese, P.~Nath, B.~D.~Nelson, G.~Peim and ,
  Phys.\ Rev.\ D {\bf 84}, 015007 (2011).



\bibitem{SS} 
  B.~C.~Allanach,
  Comput.\ Phys.\ Commun.\  {\bf 143}, 305 (2002).
(V3.3.5/3.3.6/3.37)

\bibitem{micrOMEGAs} 
  G.~Belanger, {\it et. al},
  Comput.\ Phys.\ Commun.\  {\bf 182}, 842 (2011).
(V2.4.5)


  
  


\bibitem{Ackermann:2011wa} 
  M.~Ackermann {\it et al.}  [Fermi-LAT],
  Phys.\ Rev.\ Lett.\  {\bf 107}, 241302 (2011).
 



  


\bibitem{GeringerSameth:2011iw} 
  A.~Geringer-Sameth and S.~M.~Koushiappas,
  Phys.\ Rev.\ Lett.\  {\bf 107}, 241303 (2011).
  

  

\bibitem{R} 
  R.~C.~Cotta, A.~Drlica-Wagner, S.~Murgia, E.~D.~Bloom, J.~L.~Hewett and T.~G.~Rizzo,
  JCAP {\bf 1204}, 016 (2012).
  
\bibitem{Hooper:2012sr} 
  D.~Hooper, C.~Kelso and F.~S.~Queiroz,
  arXiv:1209.3015 [astro-ph.HE].
   
   
\bibitem{DS} 
  P.~Gondolo, J.~Edsjo, P.~Ullio, L.~Bergstrom, M.~Schelke and E.~A.~Baltz,
  JCAP {\bf 0407}, 008 (2004).

 \bibitem{B} 
  M.~Ackermann {\it et al.}  [FERMI-LAT],
  Phys.\ Rev.\ D {\bf 86}, 022002 (2012)
  A.~A.~Abdo, i {\it et al.},
  Phys.\ Rev.\ Lett.\  {\bf 104}, 091302 (2010).
  
  
  
    \bibitem{A}
\url{http://fermi.gsfc.nasa.gov/science/mtgs/symposia/2012}
     
 
   
\bibitem{brem} 
  T.~Bringmann, L.~Bergstrom and J.~Edsjo,
  JHEP {\bf 0801}, 049 (2008).
  \bibitem{brem2} 
  M.~Cannoni, M.~E.~Gomez, M.~A.~Sanchez-Conde, F.~Prada and O.~Panella,
  Phys.\ Rev.\ D {\bf 81}, 107303 (2010).
  
  
  
\bibitem{hoop} 
  M.~S.~Carena, D.~Hooper and A.~Vallinotto,
  Phys.\ Rev.\ D {\bf 75}, 055010 (2007).

\bibitem{Savage} 
  J.~R.~Ellis, K.~A.~Olive and C.~Savage,
  Phys.\ Rev.\ D {\bf 77}, 065026 (2008).

\bibitem{PS} 
  S.~Amsel, K.~Freese and P.~Sandick,
  JHEP {\bf 1111}, 110 (2011);
  P.~Sandick,
  arXiv:1210.5214 [hep-ph].
  
  \bibitem{CD} 
   D.~Feldman, Z.~Liu and P.~Nath,
  Phys.\ Rev.\ D {\bf 81}, 095009 (2010);
 S.~Akula, D.~Feldman, Z.~Liu, P.~Nath and G.~Peim,
  Mod.\ Phys.\ Lett.\ A {\bf 26}, 1521 (2011).
    
    
   \bibitem{xenon1T} 
  E.~Aprile [XENON1T Collaboration],
  arXiv:1206.6288 
  
  
  \bibitem{LUX}
    S.~Fiorucci {\it et al.} [LUX],
  arXiv:1301.6942 [astro-ph.IM].
  
\bibitem{supercdms}
B.~Cabrera
 SuperCDMS Development Project  (2005).
   
    
\bibitem{gauginosLHC} 
  S.~Chatrchyan {\it et al.}  [CMS ],
  JHEP {\bf 1211}, 147 (2012).
  G.~Aad {\it et al.}  [ATLAS Collaboration],
  Phys.\ Lett.\ B {\bf 718}, 841 (2013).
  G.~Aad {\it et al.}  [ATLAS],
  Phys.\ Lett.\ B {\bf 718}, 879 (2013).
  
\bibitem{deJong} 
  P.~de Jong,
  arXiv:1211.3887 [hep-ex].
  
\bibitem{CLIC} 
  L.~Linssen, A.~Miyamoto, M.~Stanitzki and H.~Weerts,
  arXiv:1202.5940 [physics.ins-det];
see chapter 1, editors  D.~Feldman,  G.~Giudice and J. Wells, and
{ \url{ http://particle.physics.ucdavis.edu/seminars/data/media/2012/feb/wells.pdf}}

\bibitem{ilc}
 G.~Aarons {\it et al.}  [ILC Collaboration],
Physics At The ILC,''
arXiv:0709.1893 [hep-ph];
 G.~Weiglein {\it et al.}  [LHC/LC Study Group Collaboration],
 Phys.\ Rept.\  {\bf 426}, 47 (2006).

\bibitem{review} 
  P.~Nath, B.~D.~Nelson, H.~Davoudiasl, B.~Dutta, D.~Feldman, Z.~Liu, T.~Han and P.~Langacker {\it et al.},
  Nucl.\ Phys.\ Proc.\ Suppl.\  {\bf 200-202}, 185 (2010).
  
\bibitem{BBLT} 
  H.~Baer, V.~Barger, A.~Lessa and X.~Tata,
  Phys.\ Rev.\ D {\bf 86}, 117701 (2012).


\end{thebibliography}
\end{document}